# Distortion-Free High-Resolution PROPELLER-DWI: Technical Advances and Initial Demonstration for the Prostate

Jingjia Chen[1,2], Kun Zhou[3], Haoyang Pei[1,2,4], Angela Tong[1,2], Sherif Elsherif[1,2], Luke Ginocchio[1,2], Mahesh Bharath Keerthivasan[5], Qiuting Wen[6], Hersh Chandarana[1,2], Li Feng[1,2]

1. Bernard and Irene Schwartz Center for Biomedical Imaging, Department of Radiology, New York University Grossman School of Medicine, New York, New York, USA

2. Center for Advanced Imaging Innovation and Research (CAI$^2$R), Department of Radiology, New York University Grossman School of Medicine, New York, New York, USA

3. Siemens Shenzhen Magnetic Resonance Ltd., Shenzhen, China

4. Department of Electrical and Computer Engineering, New York University Tandon School of Engineering, New York, USA

5. Siemens Medical Solutions, New York, New York, USA

6. Department of Radiology and Imaging Sciences, Indiana University School of Medicine, Indianapolis, Indiana, USA

**Grant support**: NIH P41EB017183, R01EB030549, K99EB039095

Word Count: 5080

Address correspondence to:

Jingjia Chen

Department of Radiology

New York University Grossman School of Medicine

660 1st Ave, 6th Fl., New York, NY, USA 10016

Telephone: +1 212-263-0182

Email: Jingjia.Chen@nyulangone.org

## Abstract

**Purpose:** To develop and evaluate an integrated acquisition and reconstruction framework for distortion-free, high-resolution prostate diffusion-weighted imaging (DWI).

**Methods:** PROPELLER-DWI combines pulsed-gradient spin-echo diffusion preparation, an interleaved gradient-and-spin-echo PROPELLER acquisition with golden-angle blade rotation, and self-supervised joint reconstruction and denoising. It was compared with conventional NUFFT reconstruction and MPPCA denoising in a NIST/QIBA diffusion phantom, 25 volunteers, and 4 clinical patients undergoing 3T prostate MRI across three spatial-resolution settings and two b-values (50 and 1000 s/mm$^2$). Reader studies compared geometric fidelity, image quality, and overall preference with clinical single-shot EPI-DWI.

**Results:** PROPELLER-DWI preserved prostate geometry in cases where EPI-DWI showed susceptibility-induced distortion, signal loss, and metal-related artifacts. Self-supervised reconstruction improved noise suppression, with the greatest benefit at higher spatial resolution. Across all resolutions, PROPELLER-DWI had significantly higher geometric fidelity than EPI-DWI, and radiologist preference increased with the severity of EPI distortion. ADC measurements remained consistent across resolutions and comparable to clinical EPI-DWI, whereas EPI-DWI ADC varied with increasing resolution. Retrospective reduced-shot reconstruction also supported shorter scan times.

**Conclusion:** PROPELLER-DWI enables distortion-free, high-resolution prostate DWI while preserving quantitative diffusion measurements, providing a practical alternative for technically challenging prostate MRI and applications requiring improved geometric fidelity.

## Introduction

Diffusion-weighted imaging (DWI) is an essential component of multiparametric prostate MRI and plays a key role in the detection, localization, and assessment of clinically significant prostate cancer[1,2]. Within the Prostate Imaging Reporting and Data System (PI-RADS) framework, DWI and DWI-derived apparent diffusion coefficient (ADC) maps are central to lesion assessment, particularly for lesions in the peripheral zone, where the majority of prostate cancer occurs[2–6]. Current clinical prostate DWI is predominantly acquired using single-shot echo-planar imaging (EPI), which enables rapid k-space traversal following diffusion encoding to achieve a clinically feasible scan time. However, the long readout in single-shot EPI is highly sensitive to off-resonance effects. B0 inhomogeneity causes phase accumulation during the readout, resulting in geometric distortion, signal pile-up, and signal loss, which are particularly pronounced in prostate DWI, where susceptibility variations from rectal gas, bowel contents, postoperative changes, and metallic implants can produce severe geometric warping.

These limitations also restrict the achievable spatial resolution for single-shot EPI-based prostate DWI. High spatial resolution has the potential to improve lesion conspicuity and image quality. For example, prior studies[7,8] have reported improved image quality, lesion conspicuity, and prediction of adverse biopsy histology when high-resolution DWI was incorporated into MRI compared with standard DWI. However, achieving high spatial resolution with conventional single-shot EPI-DWI remains technically challenging. Beyond the inherent SNR loss from smaller voxels, the larger readout matrix lengthens the echo spacing within gradient hardware limits, slowing the traversal of k-space along the phase-encoding direction and increasing off-resonance-induced geometric distortion, while the longer echo train prolongs TE and further reduces SNR.

Several strategies have been developed to mitigate geometric distortion in EPI-based prostate DWI. In-plane parallel imaging[9–11], reduced-FOV excitation[12,13], and multi-shot EPI approaches[14–19] all improve geometric fidelity by reducing the number of acquired phase-encoding lines, shortening the effective EPI readout, or decreasing the effective echo spacing. However, because these methods remain fundamentally EPI-based, residual off-resonance distortion may persist in challenging pelvic imaging scenarios, particularly near severe susceptibility-induced B0 inhomogeneity from rectal gas, bowel contents, postoperative changes, or metallic implants. In addition, multi-shot EPI approaches can increase scan time as additional shots are required. More advanced time-resolved EPI techniques, such as echo-planar time-resolved imaging[20–22], alleviate distortion by resolving temporal signal evolution during the EPI readout rather than allowing off-resonance-induced phase accrual to manifest as spatial displacement along the phase-encoding direction. However, these methods require model-based spatiotemporal reconstruction of echo-time-resolved image series with increased reconstruction complexity and burden.

Beyond EPI-based strategies, fast spin echo (FSE) approaches offer a promising alternative that can effectively eliminate susceptibility-induced geometric distortion in DWI. In FSE-based readouts, repeated refocusing pulses reverse the phase dispersion caused by static B0 inhomogeneity, and can therefore avoid off-resonance-induced geometric distortion. However, FSE-based DWI is challenged by motion-induced phase variation during diffusion encoding, which disrupts the Carr-Purcell-Meiboom-Gill (CPMG) condition[23] and results in unstable echo pathways and signal loss. Strategies for managing non-CPMG signal often increase sequence and reconstruction complexity or reduce SNR efficiency[24–27]. FSE-based DWI is also limited by T2 blurring associated with long echo trains and prolonged scan time when multi-shot acquisitions are used[18,28–30].

In this work, we propose an integrated framework for high-resolution, distortion-free prostate DWI. The proposed framework combines pulsed-gradient spin-echo (PGSE) diffusion encoding and phase-insensitive gradient-and-spin-echo acquisition with PROPELLER sampling, and self-supervised joint reconstruction and denoising to achieve high-resolution distortion-free DWI while mitigating SNR limitations. We evaluated the proposed framework in phantom and human subjects by comparing it with clinically used single-shot EPI-DWI across multiple spatial resolutions in terms of geometric fidelity, image quality, radiologist assessment, ADC quantification, and retrospective reduced-shot reconstruction. For simplicity, this integrated framework is referred to as PROPELLER-DWI throughout the remainder of this paper.

## Methods

### PROPELLER-DWI Acquisition Design

PROPELLER-DWI was implemented as a multi-shot gradient-and-spin-echo (GRASE) PROPELLER acquisition with golden-angle rotation between shots (**Figure 1**). Diffusion encoding was performed using a conventional PGSE module preceding the readout. Because physiological motion during diffusion preparation can introduce phase dispersion that violates the CPMG condition for the FSE train, a phase-insensitive preparation was applied[24]. Specifically, following the PGSE diffusion-encoding module, a dephasing gradient was applied followed by a 90° pulse, which rotated the non-CPMG component out of the transverse plane and suppressed incoherent signal contributions. During the subsequent refocusing echo train, matched rephasing gradients were applied to preserve the CPMG-compliant signal pathway. Each shot then acquired a GRASE train[31] following a PROPELLER sampling pattern[32] (**Figure 1a**). The spin-echo train length, together with the acceleration factor, determined the blade width by setting the number of k-space lines acquired per blade, whereas the number of gradient echoes, $N_{\mathrm{GE}}$, determined the number of blades acquired within each shot. Instead of repeatedly sampling an identical trajectory to achieve multiple averages as in clinical EPI-DWI, PROPELLER-DWI acquired multiple shots with distinct blade orientations for each image. For a shot containing $N_{\mathrm{GE}}$ uniformly spaced blades, the multi-blade pattern has a rotational symmetry interval of $180°/N_{\mathrm{GE}}$ (**Figure 1b**). The multi-blade pattern was then rotated between consecutive shots by a golden-angle increment within this interval, $\Delta\theta = (180°/N_{\mathrm{GE}})/\phi$, where $\phi = 1.618$ denotes the golden ratio[33]. This sampling strategy provides complementary k-space coverage across consecutive shots, which accumulates over different diffusion-encoding directions and b-values, as illustrated in **Figure 1c**.

### Self-Supervised Joint Reconstruction and Denoising

An unrolled network was trained using a self-supervised learning scheme to jointly reconstruct and denoise PROPELLER-DWI from undersampled multicoil k-space data (**Figure 2**). The self-supervised learning followed the k-space splitting strategy[34–36] as described in a previous work[37], integrating an additional blade-wise phase-correction step for PROPELLER-DWI. Blade-wise phase correction was performed before k-space splitting.[32] For each blade, the k-space data were multiplied by a Gaussian low-pass window. The filtered data were inverse Fourier transformed to obtain a complex image, and its low-pass filtered phase $\phi_b(\boldsymbol{r})$,was removed from the corresponding unfiltered full-resolution blade image by multiplication with $e^{-i\phi_b(r)}$, followed by Fourier transform to return to the k-space. The resulting phase-corrected blade k-space data of each b-value and diffusion-encoding direction were then grouped and used for self-supervised training.

During training, the acquired multicoil blade data were randomly divided into two complementary subsets, $y_1$ and $y_2$, using disjoint sampling trajectories $\Lambda_1$ and $\Lambda_2$. The first subset, $y_1$, was used as input to the unrolled reconstruction network, while the second subset, $y_2$, was used for self-supervised data consistency loss. The network was trained to reconstruct an image from $y_1$ by predicting data points consistent with the held-out measurements in $y_2$. The network parameters were optimized according to:

$$\theta = \arg\min_{\theta} \mathbb{E}[\mathcal{L}(F_2^H W_2 F_2 f_\theta(F_1^H W_1 y_1), F_2^H W_2 y_2)]$$

where $F_1$ and $F_2$ are nonuniform fast Fourier transform (NUFFT) operators associated with the two split trajectories, $W_1$ and $W_2$ are the corresponding density-compensation matrices, and $f_\theta$ denotes the unrolled reconstruction network. The loss function $\mathcal{L}$ was implemented as a mixed $\mathcal{L}_1$ and $\mathcal{L}_2$ image-domain loss between the network reconstruction projected onto the held-out sampling operator and the image reconstructed from the held-out data.

The reconstruction network consisted of 12 cascaded iteration blocks, each alternating data-consistency updates with learned regularization following the end-to-end variational network (VarNet) structure[38]. Coil sensitivity maps were estimated from the acquired low-resolution calibration k-space using a 4-layer U-Net-based sensitivity-map estimation module and were jointly optimized with the reconstruction network during training of b=50 s/mm$^2$ images. Because of the higher SNR at b=50 s/mm$^2$, the resulting coil-sensitivity maps were subsequently fixed and used during both training and inference for b=1000 s/mm$^2$ images. During inference, the trained network was applied to the acquired k-space data without splitting to reconstruct denoised diffusion-weighted images for each b-value and diffusion direction. Trace images and ADC maps were then generated from the reconstructed diffusion-weighted images.

**Phantom Imaging Experiments**

Diffusion MRI quality assurance was performed using a NIST/QIBA diffusion phantom (CaliberMRI, Boulder, CO, USA)[39]. The phantom consists of 13 cylindrical vials, including three water vials and multiple vials of aqueous polyvinylpyrrolidone (PVP) solutions at 10%, 20%, 30%, 40%, and 50%. These PVP solutions provide a range of well-characterized isotropic ADC values for assessing diffusion measurement accuracy, bias, and linearity. Diffusion-weighted phantom imaging was performed using EPI-DWI and PROPELLER-DWI. Before imaging, the phantom was placed in the scanner bore for approximately 2 h to reach thermal equilibrium. The temperature, measured with the phantom's internal liquid-crystal temperature indicator[40], was approximately 19 ºC. Single-shot EPI-DWI was acquired at 1 x 1 $mm^2$ and 2 x 2 $mm^2$ in-plane resolutions, whereas PROPELLER-DWI was acquired at 1 x 1 $mm^2$ in-plane resolution. The slice thickness was 3 mm for all acquisitions. Detailed imaging parameters are summarized in **Supporting Table S1**.

**In Vivo Imaging Experiments**

This HIPAA-compliant, IRB-approved study included 25 male volunteers (mean age = 41.9 $\pm$16.9 years) and 4 male patients (mean age = 69.2 $\pm$ 9.1 years) who underwent clinical prostate MRI exams. Written informed consent was obtained from all participants prior to imaging. All MRI scans were performed on a 3T MRI scanner (MAGNETOM Prisma, Siemens Healthineers). Volunteers underwent axial T2-weighted imaging followed by EPI-DWI and the prototype PROPELLER-DWI at three spatial-resolution settings. For patients, PROPELLER-DWI was acquired as an add-on sequence after completion of the non-contrast portion of the clinical prostate MRI exam. Among the four patient studies, two had PI-RADS 3 lesions identified in the clinical MRI reports. Patient scans were included as feasibility examples, whereas cross-validation, reader assessment, and statistical comparisons were performed using the volunteer cohort unless otherwise specified.

To characterize the trade-off between sampling efficiency and off-resonance robustness, PROPELLER-DWI acquisitions with different numbers of gradient echoes were first evaluated in vivo while keeping the total number of blades fixed at 36 (**Supporting Figure S1**). The 6-gradient-echo configuration was selected for all subsequent experiments based on the balance between scan time and off-resonance artifacts. This configuration resulted in an inter-shot rotation angle of $(180°/6)/1.618 = 18.5°$. Fourteen spin echoes were acquired within each TR, matching the clinical prostate DWI protocol and yielding 28 phase-encoding lines per blade with 2-fold GRAPPA[11] acceleration.

Three in-plane resolution settings were evaluated: a low-resolution protocol comparable to the clinical EPI-DWI acquisition (EPI: 2.0 x 2.0 $mm^2$; PROPELLER-DWI: 1.8 x 1.8 $mm^2$), and two higher-resolution protocols of 1.4 x 1.4 $mm^2$ and 1.25 x 1.25 $mm^2$. For each resolution, diffusion-weighted images were acquired with three orthogonal diffusion directions at b=50 and b=1000 s/$mm^2$. The standard PROPELLER-DWI acquisition used 5 shots for b=50 s/$mm^2$ and 18 shots for b=1000 s/$mm^2$, with a total acquisition time of 6 min 50 s. Detailed imaging parameters are summarized in **Supporting Table S1**. For the patient feasibility studies, only the highest-resolution PROPELLER-DWI protocol (1.25 x 1.25 $mm^2$) was acquired to minimize disruption to the clinical workflow.

**Conventional Reconstruction and MPPCA Denoising**

For conventional reconstruction, raw multicoil blade data were first reconstructed using GRAPPA to recover missing phase-encoding lines within each blade. Low-frequency image phase caused by physiological motion and gradient echo dephasing was then estimated and removed from each blade using the Gaussian-filtered phase estimate described above. Blades corresponding to the same b-value and diffusion direction were then combined and reconstructed using NUFFT[41,42]. The density-compensation weights were calculated separately for each diffusion-encoding direction using its corresponding sampling trajectory. Briefly, unit values were assigned to all sampled non-Cartesian k-space locations, the adjoint NUFFT was applied to obtain the corresponding image-space weighting, and the forward NUFFT of its reciprocal was used as the density-compensation weights. Diffusion trace images were generated from images of three directions, and ADC maps were calculated from the b=50 and b=1000 s/mm$^2$ images. As a conventional denoising comparison, Marchenko-Pastur principal component analysis (MPPCA) denoising[43,44] was applied to the raw data along the coil dimension for each blade. The denoised blade data were then reconstructed using the same GRAPPA and NUFFT pipeline used for conventional reconstruction.

**Self-Supervised Training**

The proposed self-supervised reconstruction was evaluated using five repeated train-test rounds across the 25 in vivo datasets. For each round, datasets of 5 unique subjects and one common representative subject were held out for inference and an independent model was trained using the remaining 19 subjects. This design maximized the number of subjects available for training in each round while ensuring that each held-out subject was reconstructed using a model that had not been trained on that subject. The training and testing case assignments are provided in **Supporting Table S2**.

During training, the k-space split ratio was randomly varied from 0.3-0.99 to expose the network to a broad range of undersampling levels and improve robustness. During inference, the trained model was applied to the full acquired k-space data without splitting. Models were trained using the Adam optimizer with an initial learning rate of 0.0001, batch size of 1, and 200 training epochs. The model was implemented in PyTorch (v2.0) with CUDA 11.7 and trained on a high-performance workstation equipped with an NVIDIA A100 GPU (NVIDIA, Santa Clara, CA, USA).

To assess the feasibility of scan time reduction, PROPELLER-DWI datasets were retrospectively reconstructed using subsets of the acquired shots. Full-shot reconstructions used 5 shots for b=50 s/mm$^2$ and 18 shots for b=1000 s/mm$^2$, whereas reduced-shot reconstructions used 3/12 and 2/6 shots for b=50/b=1000 s/mm$^2$, corresponding to total acquisition times of 4 min 30 s and 2 min 29 s, respectively.

**Image Quality Assessment**

Two fellowship-trained body radiologists independently evaluated image quality. For each case, spatial-resolution setting, and image type, EPI-DWI and PROPELLER-DWI images were displayed side-by-side with randomized left-right order. Readers scored each image separately for perceived overall image quality, sharpness, and artifacts using a 5-point Likert scale, with higher scores indicating better image quality. Readers also recorded a preference for each paired comparison. The quality assessment was performed separately for b=50 s/mm$^2$ images, b=1000 s/mm$^2$ images, and ADC maps.

**Geometric Fidelity Assessment**

Because geometric distortion is a distinct failure mode from general image quality, geometric fidelity was assessed in a separate reader task. The approximately 50-minute protocol introduced scan-to-scan motion and nonrigid anatomic changes, including bladder filling and bowel-related displacement, limiting reliable registration- or segmentation-based distortion quantification. Geometric fidelity was assessed using b=50 s/mm$^2$ DWI images paired with the corresponding T2-weighted image as the anatomic reference. Readers scored geometric fidelity on a 4-point ordinal scale: 1 = severe distortion, 2 = moderate distortion with substantial prostate shape change, 3 = mild distortion or mild signal pile-up, and 4 = no visible distortion, shape change, or signal pile-up. Higher scores indicated better geometric fidelity with less visible distortion.

**Statistical Analysis**

Scores from the two readers were averaged for each case before statistical analysis. Image-quality scores for each criterion were compared between EPI-DWI and PROPELLER-DWI using paired Wilcoxon signed-rank tests for each resolution and image type (b=50 s/mm$^2$, b=1000 s/mm$^2$, and ADC). Geometric fidelity scores were compared between EPI-DWI and PROPELLER-DWI using paired Wilcoxon signed-rank tests within each resolution setting. Benjamini-Hochberg correction was applied separately to image-quality and geometric fidelity comparisons. To assess whether preference depended on EPI distortion, averaged EPI distortion scores were grouped into bins, and the method preferred by both readers within each bin was evaluated using exact binomial tests. For ADC analysis, four circular ROIs were manually placed on ADC maps, including left and right ROIs in the peripheral zone and transition zone. ADC values were averaged for all subjects. Left-right ADC differences were assessed using a paired t-test. Resolution dependence was assessed for each sequence-zone combination using linear mixed-effects models with subject-specific random intercepts, comparing ADC ~ resolution + (1 | subject) against a null model ADC ~ 1 + (1 | subject) by a likelihood-ratio test. Holm correction was applied across the four comparisons and adjusted $P < 0.05$ was considered significant. Inter-reader agreement is reported using weighted kappa and within $\pm 1$ agreement.

## Results

In the phantom experiment, shown in **Figure 3**, the metal bolt (red box region) and adjacent air bubble (green box region) caused geometric distortion in images acquired using single-shot EPI-DWI, which was worsened at higher spatial resolution. Under similar scan times of ~5 min (**Supporting Table S1**), the PROPELLER-DWI images maintained correct geometry despite a slight decrease in the apparent SNR.

**Figure 4** compares reconstruction strategies for PROPELLER-DWI at low and high spatial resolutions. MPPCA denoising visibly reduced noise at the low-resolution setting but was less effective at higher spatial resolution, where SNR was lower and fine anatomic detail was more vulnerable to noise. The self-supervised reconstruction provided the best overall image quality across b=50 s/mm$^2$, b=1000 s/mm$^2$, and ADC images, with improved noise suppression and preservation of anatomic detail. For all three resolution settings of 1.8 x 1.8 mm$^2$, 1.4 x 1.4 mm$^2$, 1.25 x 1.25 mm$^2$, the inference times for the PROPELLER-DWI datasets were 9.94 ± 1.49 seconds, 14.27 ± 1.48 seconds, and 16.17 ± 1.61 seconds per case, respectively. During each round of training, dataset of one common case was held out to assess the model-to-model consistency (**Supporting Table S2**) by reconstructing the same dataset with all five independently trained models. Across all three spatial resolutions, structural similarity index metric (SSIM) remained high for b=50 s/mm$^2$ images (0.967-0.985) and b=1000 s/mm$^2$ images (0.948-0.976), with 5x-scaled difference maps showing only model-dependent variation in the low signal background region (**Supporting Figure S2**).

**Figures 5 and 6** show DWI images of two representative subjects. In a subject with excessive rectal gas, EPI-DWI showed distortion and signal degradation adjacent to the gas-filled rectum, particularly at higher resolution (**Figure 5**). PROPELLER-DWI preserved prostate geometry and maintained interpretable b=50 s/mm$^2$, b=1000 s/mm$^2$, and ADC images across the tested resolution settings. **Figure 6** demonstrates the performance of PROPELLER-DWI in a subject with fiducial markers after radiotherapy for prostate cancer. The metallic fiducial markers, visible in CT images, produced prominent distortion and artifact in images acquired with clinical single-shot EPI-DWI, which was worsened in high b-value and high-resolution images. In contrast, PROPELLER-DWI substantially reduced the metal-related artifact and preserved anatomic structures within the prostate, especially at high b-value and high-resolution settings. Two cases with PI-RADS 3 transition zone lesions from the clinical add-on scan experiments are shown in **Supporting Figure S3**. Although no conspicuous gas-related artifact was present on the lesion-containing slices, lesion conspicuity and ADC appearance on PROPELLER-DWI were concordant with the clinical EPI-DWI findings, supporting the feasibility of PROPELLER-DWI for lesion depiction in clinically relevant cases.

ADC analysis is summarized in **Figure 7**. In the phantom, PROPELLER-DWI ADC measurements agreed with ground-truth values and with the EPI-DWI ADC values. In vivo, the single-shot EPI-DWI showed significant resolution-dependent ADC trends (adjusted $p = 5.59 \times 10^{-8}$ for peripheral zone and adjusted $p = 2.39 \times 10^{-20}$ for transition zone), likely caused by reduced SNR in the b=1000 s/mm$^2$ images and signal loss from the longer TE at higher resolution settings, consistent with the signal degradation in the high-resolution images shown in **Figures 5 and 6**. The signal loss can bias ADC estimation downward, particularly in the transition zone. In contrast, PROPELLER-DWI showed more consistent ADC values across resolution settings. No significant left-right differences in ADC measurements were observed (**Supporting Figure S4**).

Averaged image quality scores across the two readers comparing EPI-DWI and PROPELLER-DWI for images of b=50 s/mm$^2$, b=1000 s/mm$^2$ and ADC are shown in **Figure 8**. For image quality scores, weighted kappa values ranged from 0.06 to 0.22; however, the agreement within ±1 score was 84.0-87.6% for PROPELLER-DWI and 60.4-65.8% for EPI-DWI, which reflects the known limitation of weighted kappa with narrow or skewed score distributions[45,46]. For overall image quality, sharpness, and artifact level, PROPELLER-DWI generally received higher reader scores than EPI-DWI across the three spatial-resolution settings. The improvement was most consistent for artifact scores, where PROPELLER-DWI was rated significantly higher than EPI-DWI for all image types and resolutions. For overall image quality and sharpness, PROPELLER-DWI was also scored significantly higher in most comparisons, with nonsignificant differences only for low-resolution b=1000 s/mm$^2$ images and for the highest-resolution ADC maps.

Geometric fidelity scores averaged across the two readers for EPI-DWI and PROPELLER-DWI are summarized in **Supporting Figure S5**. PROPELLER-DWI demonstrated significantly higher geometric fidelity than EPI-DWI across all three spatial-resolution settings, indicating reduced visible distortion and prostate shape deformation. Inter-reader agreement between the two readers for geometric fidelity scoring showed a weighted kappa of 0.38, with 76.7% agreeing within $\pm 1$ score. **Figure 9** illustrates the reader preference stratified by the average EPI-DWI geometric fidelity score. Readers significantly favored PROPELLER-DWI when EPI-DWI geometric fidelity was poor. In the intermediate geometric-fidelity range, no significant preference was observed, with a larger proportion of cases showing no preference by both readers. The corresponding resolution-stratified analysis is shown in **Supporting Figure S6**.

**Figure 10** demonstrates the feasibility of reducing PROPELLER-DWI scan time by reconstructing fewer acquired shots at the 1.25 x 1.25 mm$^2$ resolution. The reduced-shot reconstructions remained visually similar to the full-shot reference, with progressively increasing noise as the number of shots decreased. Difference maps relative to the full-shot reconstruction showed small ADC differences for both the 4 min 30 s and 2 min 29 s reconstructions, supporting preservation of ADC values in this representative case. The 2 min 29 s reconstruction showed greater noise and larger difference-map changes, particularly for b=1000 s/ mm$^2$ and ADC but preserved overall prostate anatomy and ADC contrast.

## Discussion

This study demonstrates the feasibility of high-resolution, distortion-free prostate DWI using an integrated framework combining GRASE PROPELLER acquisition and self-supervised reconstruction. The main finding is that PROPELLER-DWI preserved prostate geometry in settings where conventional single-shot EPI-DWI exhibited susceptibility-induced distortion, signal loss, or metal-related artifacts, while supporting higher spatial resolution and consistent ADC measurements across resolution settings. This framework addresses two major technical challenges in prostate DWI: susceptibility-induced distortion arising from the pelvic region and the SNR penalty associated with FSE-based DWI, particularly at high spatial resolution. These findings suggest that PROPELLER-DWI may serve as a practical complementary acquisition for technically challenging prostate DWI exams, particularly in the presence of rectal gas, postoperative changes, metallic implants, or when high spatial resolution is desired.

The distortion resistance of PROPELLER-DWI primarily arises from the FSE component of its readout. In EPI-DWI, successive phase-encoding lines are acquired at progressively later gradient echoes and therefore the off-resonance accumulates a phase ramp along the phase encoding direction, producing a spatial displacement given by $\Delta y = \Delta f \times \mathrm{ESP}_{\mathrm{eff}} \times \mathrm{FOV}_{PE}$, where $\mathrm{ESP}_{\mathrm{eff}}$ is the effective echo spacing and $\mathrm{FOV}_{\mathrm{PE}}$ is the FOV along the phase-encoding direction. In PROPELLER-DWI, repeated spin-echo refocusing prevents such unidirectional phase accumulation along the phase-encoding direction, thereby substantially reducing off-resonance-induced geometric distortion. Conventional FSE-based DWI can reduce susceptibility-related geometric distortion but is often limited by T2 blurring and prolonged scan time. Single-shot FSE-based variants, such as HASTE-DWI[47], shorten scan time but remain constrained by T2 blurring and SNR limitations. In comparison, the GRASE readout used in this work improves acquisition efficiency and reduces the number of refocusing pulses compared to FSE-only approaches, thereby lowering the specific absorption rate (SAR) while preserving geometric fidelity. However, this design also comes with a trade-off, where the gradient-echo component introduces residual off-resonance sensitivity. Because individual blades are acquired at different gradient-echo times relative to the refocused spin echo, they carry different gradient-echo-time-dependent T2* and off-resonance weighting. Blades acquired at later gradient-echo times are therefore more susceptible to off-resonance dephasing and can exhibit greater blurring or blooming near strong susceptibility sources. Accordingly, although a larger number of gradient echoes was feasible in many healthy volunteers, a conservative six-gradient-echo configuration was selected to balance acquisition efficiency with robustness to off-resonance artifacts, particularly in the presence of metallic implants. Advanced EPI-based approaches, including reduced-FOV and multishot EPI, can further mitigate susceptibility-induced distortion and achieve shorter TE than the high-resolution single-shot EPI protocols evaluated here. At our institution, reduced-FOV ZOOMit is routinely used as an alternative when conventional EPI-DWI for the prostate is substantially degraded by susceptibility effects. As shown in Supporting Figure S7, in a subject with a metallic implant within the prostate, ZOOMit reduced distortion compared with conventional EPI-DWI, but substantial residual artifact remained, whereas PROPELLER-DWI better preserved the prostate anatomy. Systematic evaluation in a larger cohort of patients with metallic implants is warranted to further establish the clinical utility of PROPELLER-DWI in this setting.

Beyond reducing geometric distortion, PROPELLER-DWI also enables high-resolution prostate imaging within a clinically feasible scan time. Although high spatial resolution has the potential to improve lesion conspicuity, lesion delineation, and anatomic localization,[7,8] achieving high-resolution prostate DWI with conventional single-shot EPI remains technically challenging. In current clinical prostate MRI protocols, DWI is typically acquired at a relatively coarse spatial resolution of approximately 2 x 2 x 3 $mm^3$, which is substantially lower than that of T2-weighted imaging. This compromise arises because achieving higher spatial resolution with single-shot EPI leads to slower traversal of $k_y$ as acquisition speed is limited by the available gradient performance, thereby reducing the effective phase-encoding bandwidth. The associated longer TE and smaller voxel size further reduce SNR. Consequently, high-resolution EPI can become increasingly susceptible to off-resonance-induced geometric distortion and signal loss, as demonstrated in our phantom experiments. In contrast, PROPELLER-DWI maintained a consistent TE across the tested spatial resolutions, allowing higher-resolution acquisition without the additional TE penalty imposed by single-shot EPI-DWI.

The self-supervised reconstruction and denoising framework was an important component of the proposed framework. Although the proposed acquisition substantially reduces geometric distortion, the smaller voxel size required for high-resolution imaging inevitably introduces an SNR penalty. Because fully sampled, high-SNR reference datasets are difficult to acquire for prostate DWI, the self-supervised learning strategy enabled joint reconstruction and denoising directly from the acquired undersampled PROPELLER data[37]. A blade-wise low-resolution phase correction was incorporated before data splitting to compensate for phase inconsistencies arising from both the gradient-and-spin-echo readout and physiological motion during diffusion encoding. The proposed reconstruction substantially reduced image noise while preserving fine anatomic structure and provided greater improvement than conventional MPPCA denoising, particularly at higher spatial resolution. These results suggest that integrating advanced reconstruction with distortion-free acquisition is essential for achieving clinically practical high-resolution prostate DWI.

The reader study provided clinical context for these technical advantages. Radiologist preference depended strongly on the severity of EPI distortion. In cases with substantial EPI-DWI distortion, PROPELLER-DWI consistently preserved prostate anatomy and was more frequently preferred by readers. In contrast, when EPI-DWI distortion was mild, reader preference was more balanced. At conventional clinical resolution, EPI-DWI remains highly efficient in scan time and SNR. However, it struggles at higher spatial resolution, where the longer EPI readout and associated increase in TE can lead to greater susceptibility-induced distortion, signal loss, and reduced SNR. Therefore, for high-resolution settings and in technically challenging cases, PROPELLER-DWI may serve as a complementary acquisition. Additionally, the improved spatial correspondence between diffusion abnormalities and prostate anatomy may also support applications that require accurate lesion localization, such as focal therapy planning and radiation-therapy dose boosting with MR-LINAC[48,49].

The ADC results indicate that acquisition variations can impose bias in quantitative diffusion measurements. Although tissue diffusivity in principle should be independent of spatial resolution, ADC values estimated using EPI-DWI showed decreasing ADC values with increasing spatial resolution, particularly in the transition zone of the prostate. This is likely due to the reduced SNR, longer TE, and increased signal loss associated with long TE at higher EPI-DWI resolution. In particular, signal voids, especially in the transition zone, and noise-floor bias at b = 1000 s/mm$^2$ images may have artificially increased the high-to-low b-value signal ratio, leading to ADC underestimation[50,51]. These effects, to an extent, explain why conventional prostate DWI is typically acquired at relatively low spatial resolution. In contrast, PROPELLER-DWI maintained a consistent TE; therefore, ADC values were consistent across different resolutions. However, we observed that the estimated ADC values from PROPELLER-DWI images were slightly higher than those obtained with low-resolution EPI-DWI. The slightly higher in vivo ADC observed with PROPELLER-DWI may partly reflect motion-induced signal attenuation in the multi-shot acquisition, which can mimic additional diffusion and lead to apparent ADC overestimation.

Several limitations of this study should be acknowledged. First, the current implementation remains limited by SNR, particularly at high b-values and high spatial resolutions. Although the proposed self-supervised reconstruction substantially mitigated this limitation, further improvements may be achieved with larger and more diverse training datasets. In addition, approximately half of the signal is discarded by the phase-insensitive preparation used to maintain a stable spin-echo train. Future work will investigate other approaches, including optimized phase cycling and non-CPMG signal stabilization, to recover the full signal pathways and improve SNR efficiency[27,28,52]. Second, the fully sampled PROPELLER-DWI acquisition requires longer scan time than conventional single-shot EPI-DWI, although retrospective reduced-shot reconstruction demonstrated the feasibility of substantially shorter acquisitions. The present study also did not include a systematic comparison with advanced EPI-based approaches, such as reduced-FOV or multi-shot EPI, which can reduce susceptibility-induced distortion and achieve shorter TE than the high-resolution single-shot EPI protocols evaluated here. The current implementation used centric echo ordering within each blade to acquire central k-space at the highest available signal level. Because the acquisition alternates between the two sides of k-space while progressing outward, the echo time increases by approximately two echo spacings per phase-encoding step along either side of the blade, resulting in a symmetric T2-dependent low-pass weighting and associated blurring within each blade. Importantly, the blade-level T2 blurring does not translate directly into the same degree of blurring in the final PROPELLER image, as the final point-spread function is determined by the superposition of the overlapping T2-weighted k-space contributions from multiple rotated blades. Future work will investigate alternative reordering schemes, such as linear ordering, which produces a less steep and asymmetric T2-dependent weighting that may be partially averaged across rotated blade orientations, as well as model-based T2 blur correction[53] during image reconstruction. Third, compared with EPI-DWI, the gradient-and-spin-echo PROPELLER readout interleaves gradient-echo encoding with refocusing RF pulses and may reduce the readout gradient duty cycle. Future studies will be needed to investigate the practical implications of this acquisition design, including peripheral nerve stimulation and patient comfort during broader clinical implementation[54]. Finally, the current reader study focused primarily on geometric fidelity, image quality, and reader preference. Although these results demonstrate the technical feasibility of the proposed framework, larger prospective studies with PI-RADS-based diagnostic assessment and histopathologic correlation are needed to determine whether the improved geometric fidelity and higher spatial resolution translate into improved lesion detection, localization, and risk stratification.

## Conclusion

This work presents an integrated framework for distortion-free high-resolution prostate DWI by combining golden-angle rotating interleaved gradient-and-spin-echo PROPELLER acquisition with self-supervised joint reconstruction and denoising. Compared with the current clinical standard, the proposed framework enables higher spatial resolution while preserving geometric fidelity within a clinically feasible scan time, with retrospective reduced-shot reconstruction demonstrating the potential for further scan-time reduction. Radiologists preferred PROPELLER-DWI particularly in cases where EPI-DWI suffered from geometric distortion. These results support PROPELLER-DWI as a potential alternative for high-resolution quantitative and diagnostic prostate DWI, particularly in technically challenging pelvic imaging settings.

## Acknowledgment

The authors would like to thank Dr. James Pipe and Dr. Anton Becker for helpful discussion and feedback. This work was supported by the NIH NIBIB (P41EB017183, R01EB030549, K99EB039095) and was performed under the rubric of the Center for Advanced Imaging Innovation and Research ($CAI^2R$), an NIBIB Biomedical Technology Resource Center. This work was also supported by the Institute of Information & communications Technology Planning & Evaluation (IITP) grant funded by the Korean government (MSIT) (No. RS-2024-00457882, National AI Research Lab Project).

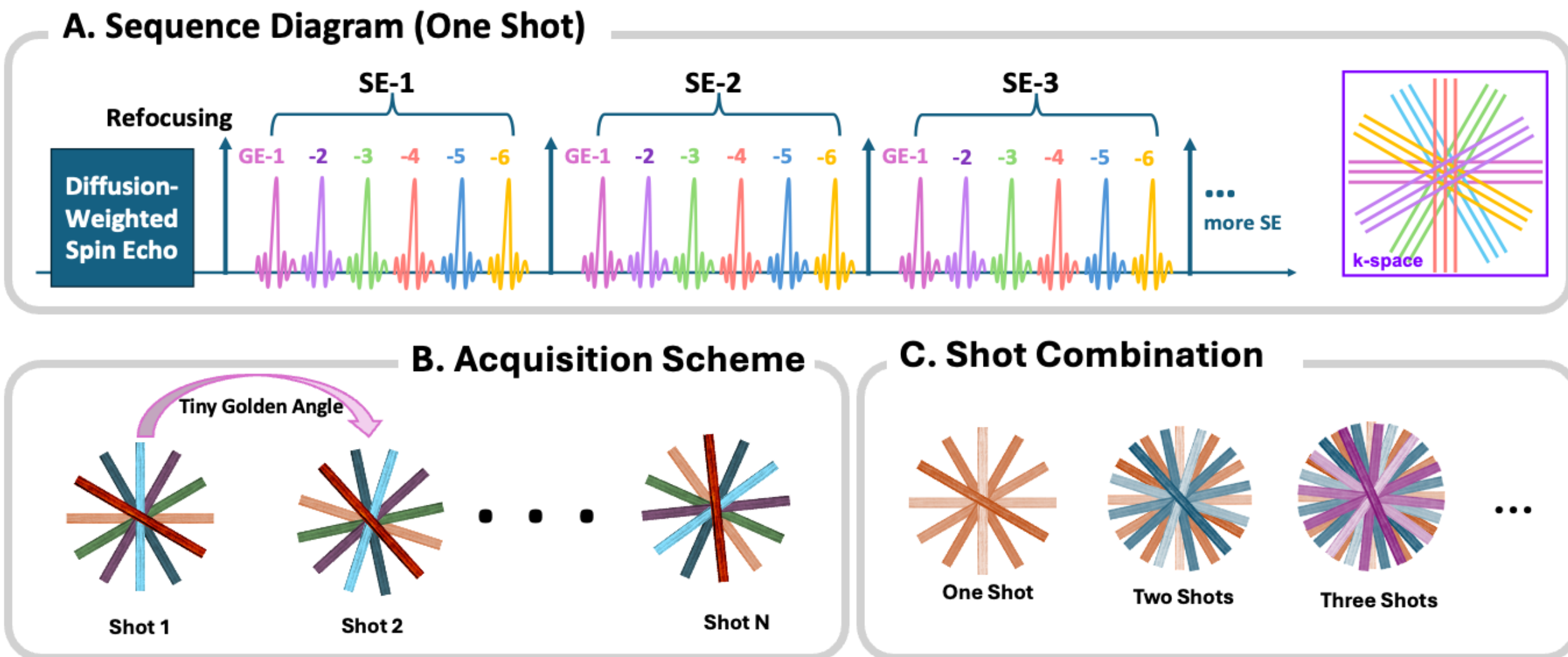

**Figure 1**. PROPELLER-DWI acquisition design. (A) Sequence diagram of the gradient-and-spin-echo PROPELLER readout after diffusion-weighted spin-echo preparation. Multiple gradient echoes are acquired around each refocusing pulse, and each spin echo contributes a group of gradient-echo blades in k-space. (B) Acquisition scheme using scaled-golden-angle rotation between consecutive shots. (C) Progressive combination of shots produces complementary k-space coverage for each diffusion weighting and diffusion-encoding direction.

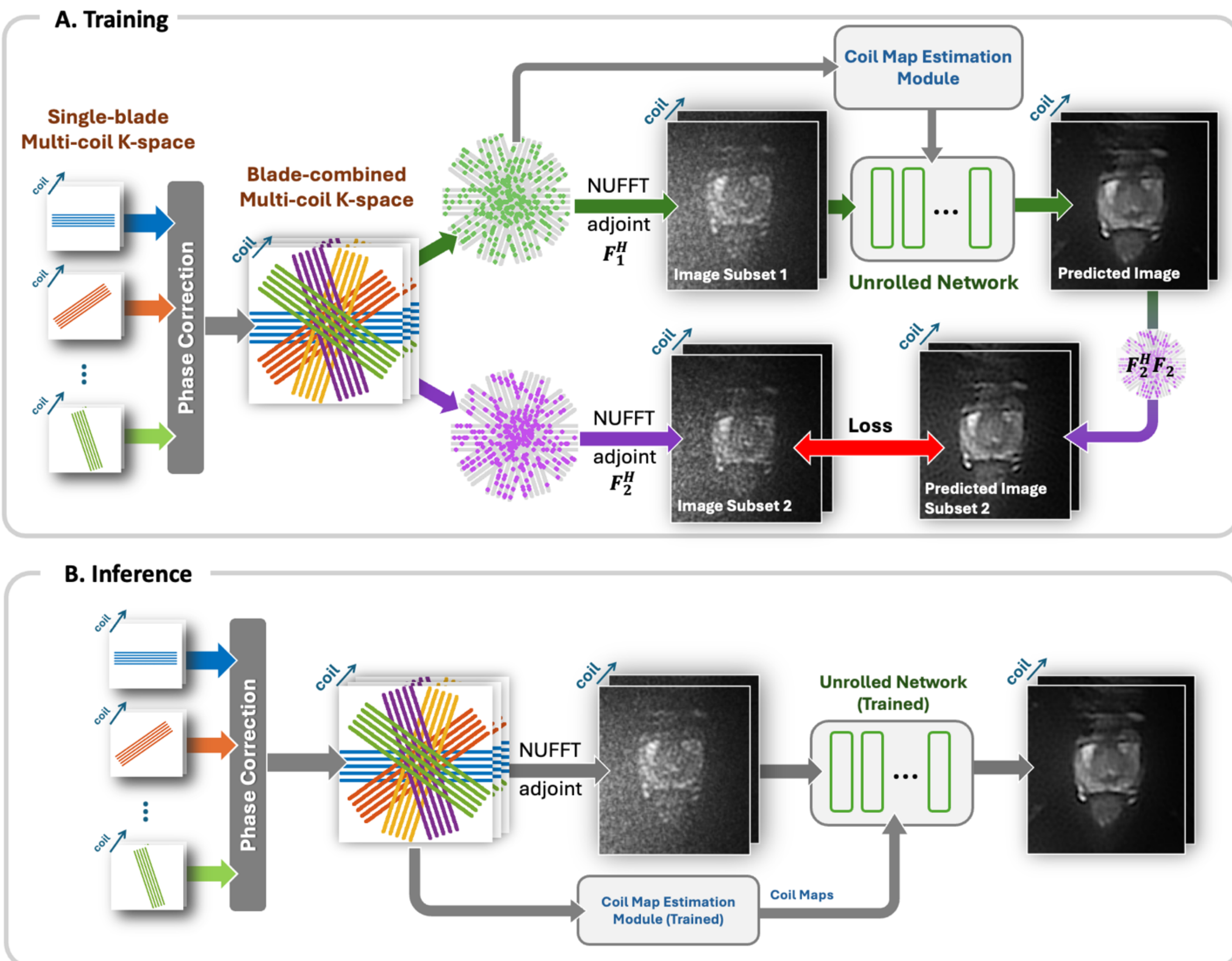

**Figure 2**. Self-supervised reconstruction and denoising framework for PROPELLER-DWI. (A) During training, single-blade multicoil k-space data undergo blade-wise phase correction and are combined into complementary k-space subsets. One subset is used as input to the unrolled reconstruction network, while the held-out subset is used for self-supervised loss calculation after projection through the corresponding NUFFT operator. Coil sensitivity maps are estimated from low-resolution data and optimized jointly with the reconstruction network. (B) During inference, the trained network and coil-map estimation module are applied to the full phase-corrected blade-combined k-space data to generate the final denoised image.

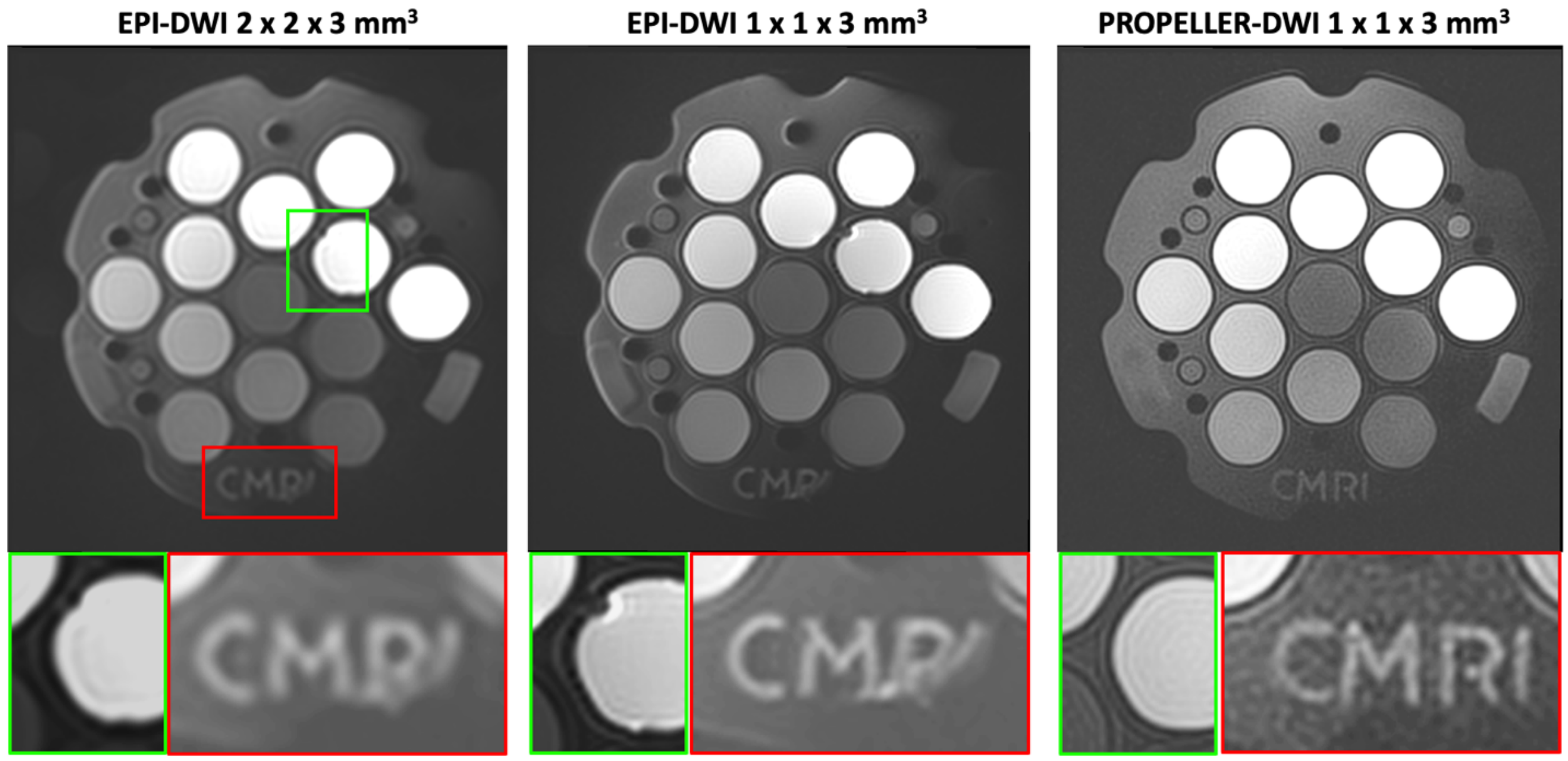

**Figure 3**. Phantom evaluation of susceptibility-related distortion. Diffusion-weighted images of a NIST/QIBA diffusion phantom acquired with EPI-DWI at 2.0 x 2.0 x 3 $mm^3$ and 1.0 x 1.0 x 3 $mm^3$ resolution, and with PROPELLER-DWI at 1.0 x 1.0 x 3 $mm^3$ resolution. Magnified regions show susceptibility-related distortion near the bolt and adjacent air bubble on EPI-DWI, with more severe distortion at higher EPI resolution. PROPELLER-DWI preserves the expected phantom geometry and reduces susceptibility-induced deformation.

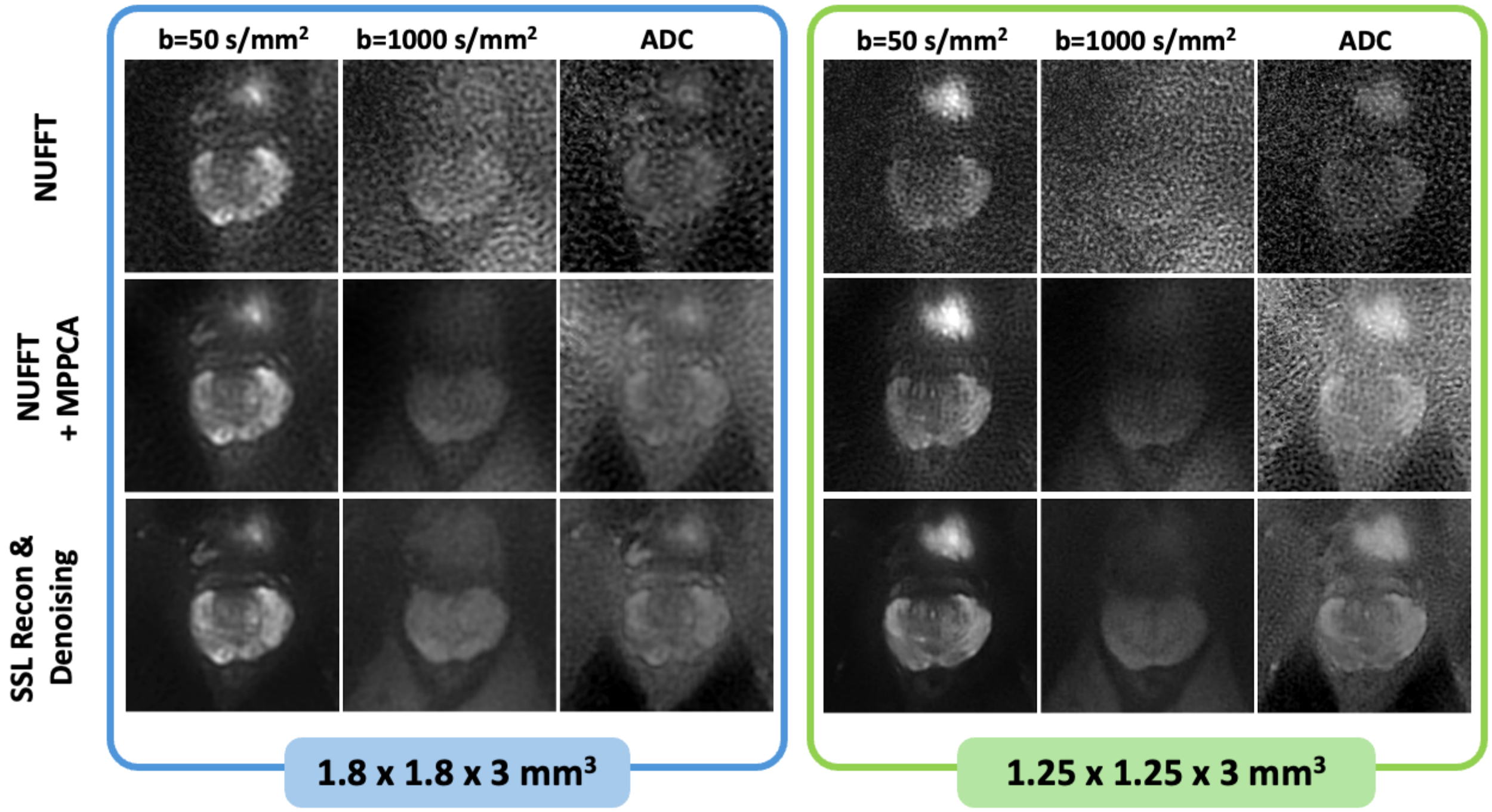

**Figure 4**. Comparison of PROPELLER-DWI reconstruction strategies at low and high spatial resolution. Representative b=50 s/mm$^2$, b=1000 s/mm$^2$, and ADC images reconstructed using conventional NUFFT, NUFFT with MPPCA denoising, and the proposed self-supervised learning (SSL) reconstruction and denoising method. Results are shown at 1.8 x 1.8 x 3 mm$^3$ and 1.25 x 1.25 x 3 mm$^3$ resolution. The self-supervised reconstruction reduces noise while preserving prostate anatomy, with greater visible benefit at higher spatial resolution.

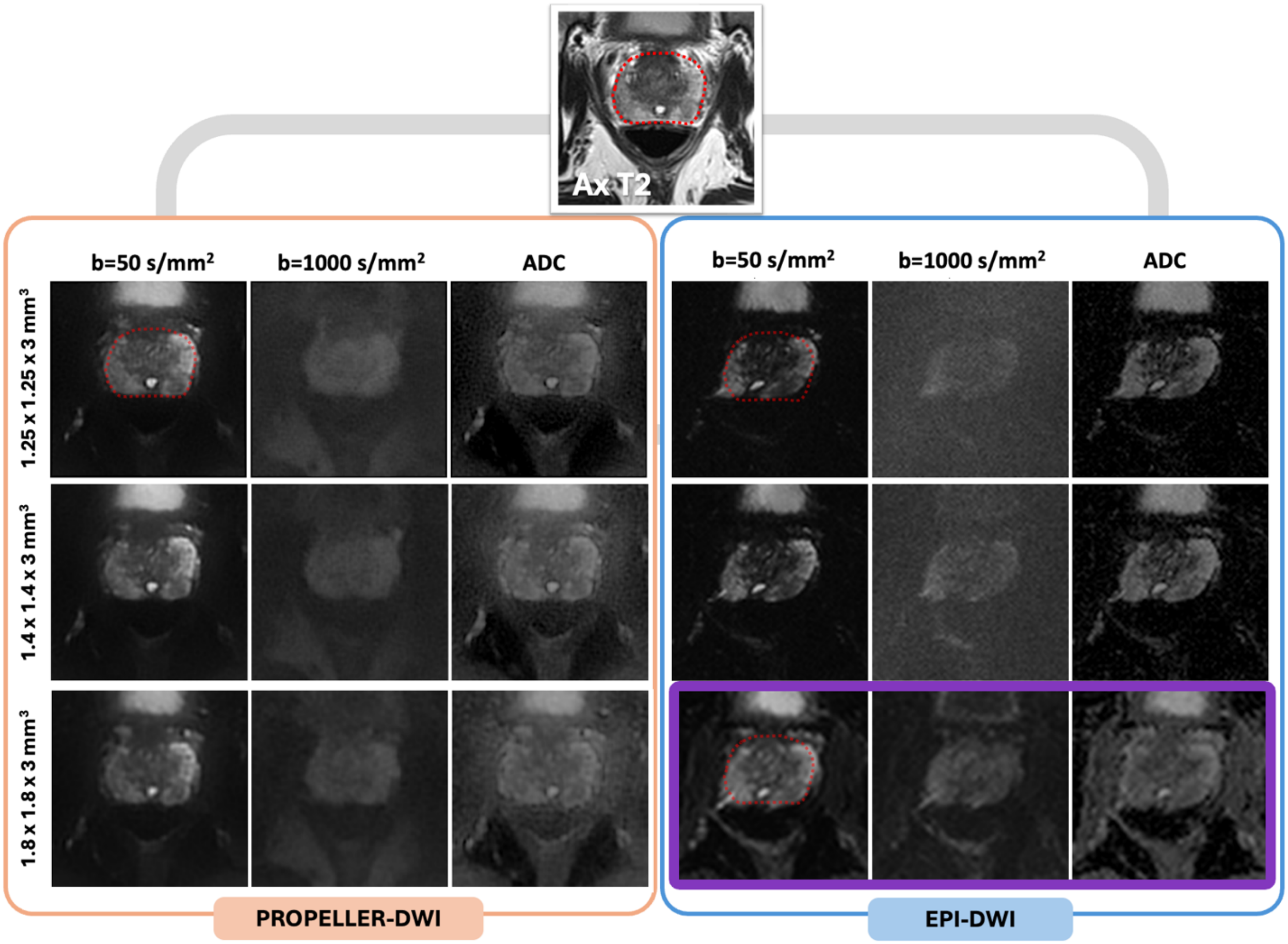

**Figure 5**. Representative in vivo comparison in a subject with rectal gas-related susceptibility artifact. Axial T2-weighted image and corresponding EPI-DWI and PROPELLER-DWI images are shown across three spatial-resolution settings. Purple box indicates the current clinical protocol. EPI-DWI demonstrates geometric distortion and signal degradation adjacent to the gas-filled rectum, particularly at higher spatial resolution. PROPELLER-DWI preserves prostate geometry and maintains interpretable b=50 s/mm$^2$, b=1000 s/mm$^2$, and ADC images across the tested resolutions.

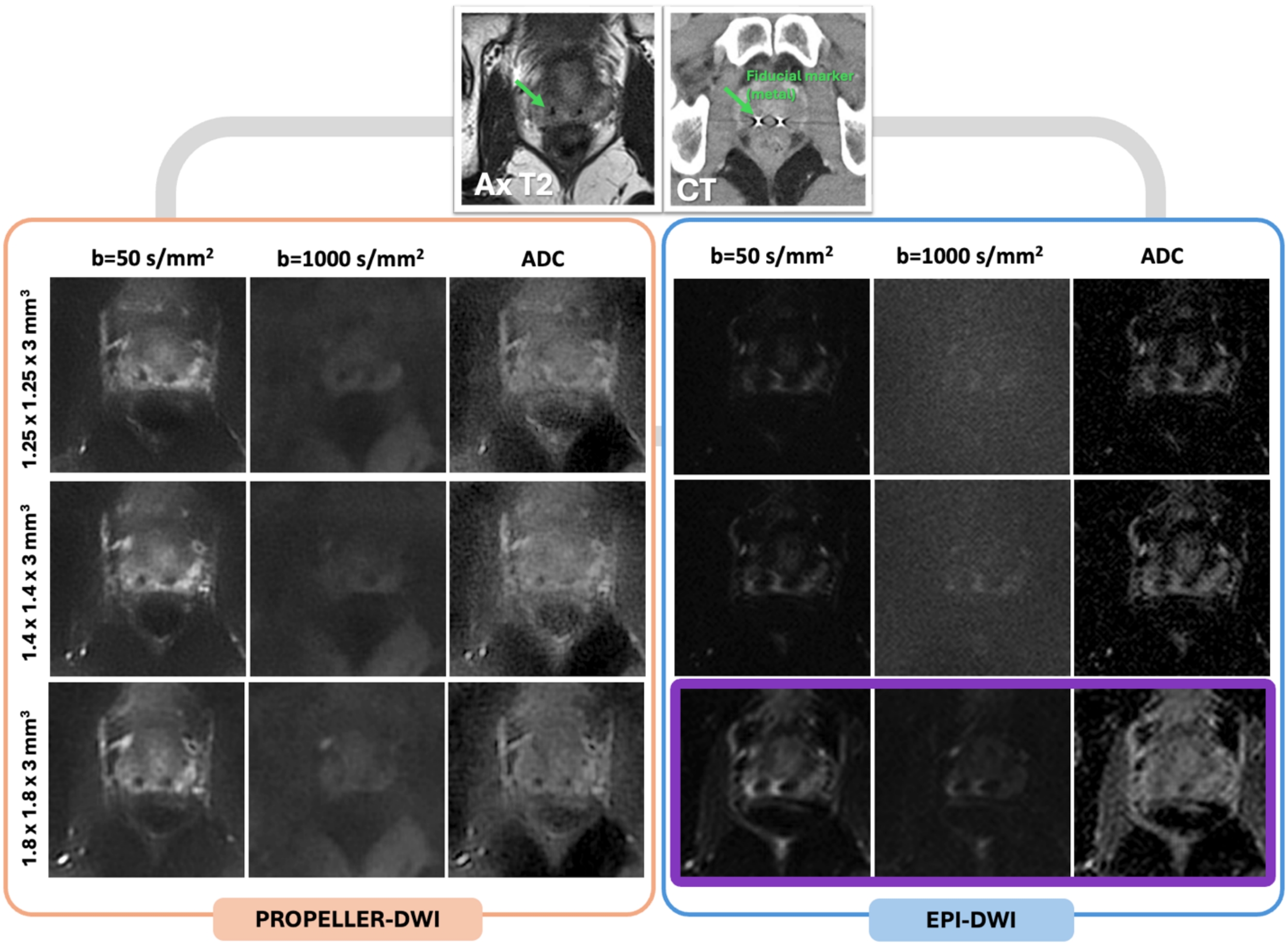

**Figure 6**. Representative in vivo comparison in a subject with metallic fiducial markers. Axial T2-weighted and CT images identify residual fiducial markers after radiotherapy. Purple box indicates the current clinical protocol. EPI-DWI shows marked metal-related distortion and signal loss, which becomes more limiting at higher spatial resolution. PROPELLER-DWI reduces metal-related artifact and preserves prostate anatomy on b=50 s/mm$^2$, b=1000 s/mm$^2$, and ADC images across the tested resolutions.

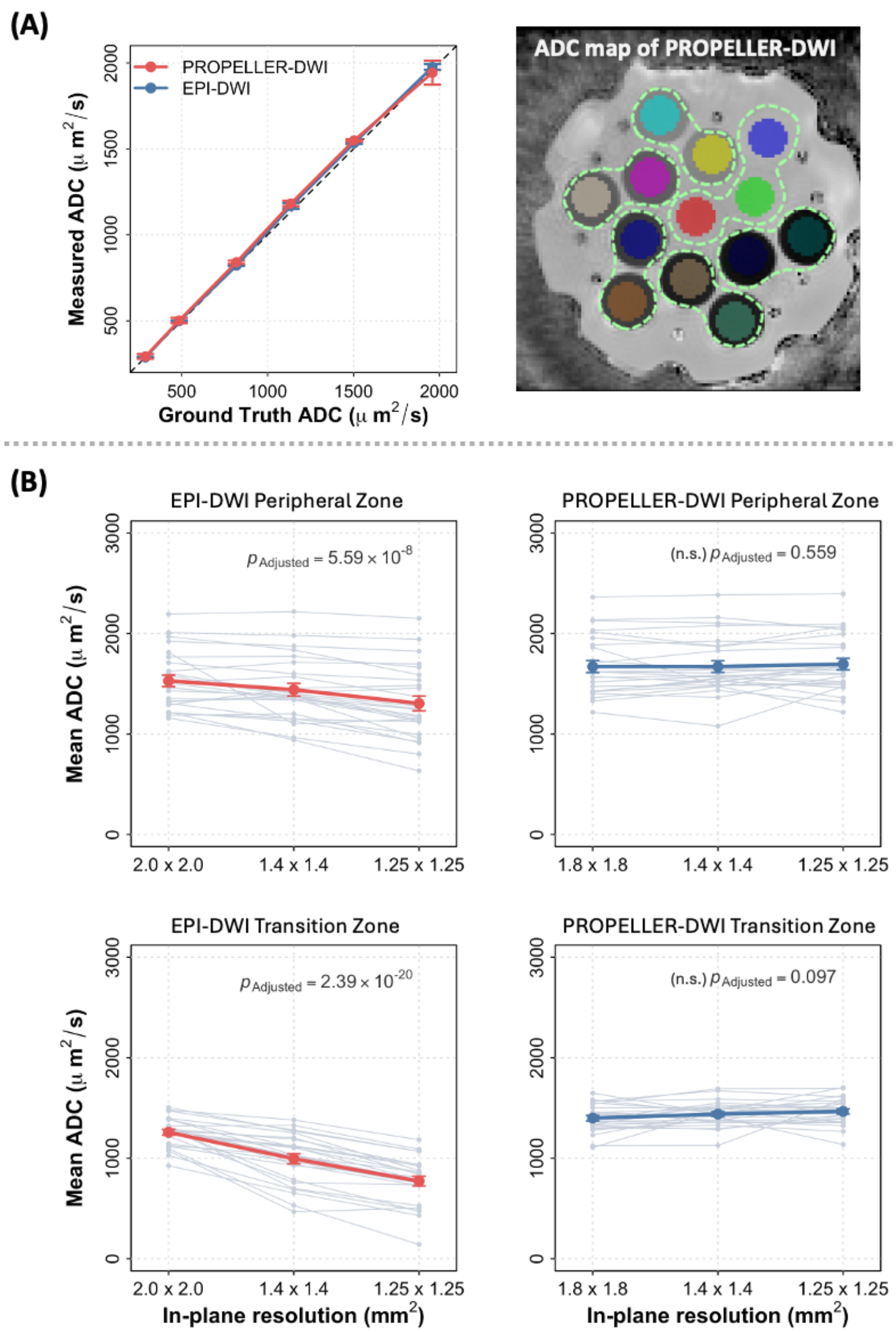

**Figure 7**. ADC analysis in phantom and in vivo experiments. (A) Phantom ADC measurements from EPI-DWI and PROPELLER-DWI compared with ground-truth ADC values in the NIST/QIBA diffusion phantom. The representative PROPELLER-DWI ADC map shows the ROI locations used for vial-wise measurement. Error bars indicate the standard deviation among vials of the same PVP concentration. (B) In vivo prostate ADC measurements in the peripheral zone and transition zone across spatial resolutions. EPI-DWI shows significant resolution-dependent ADC changes, whereas PROPELLER-DWI shows more stable ADC measurements across the tested resolutions.

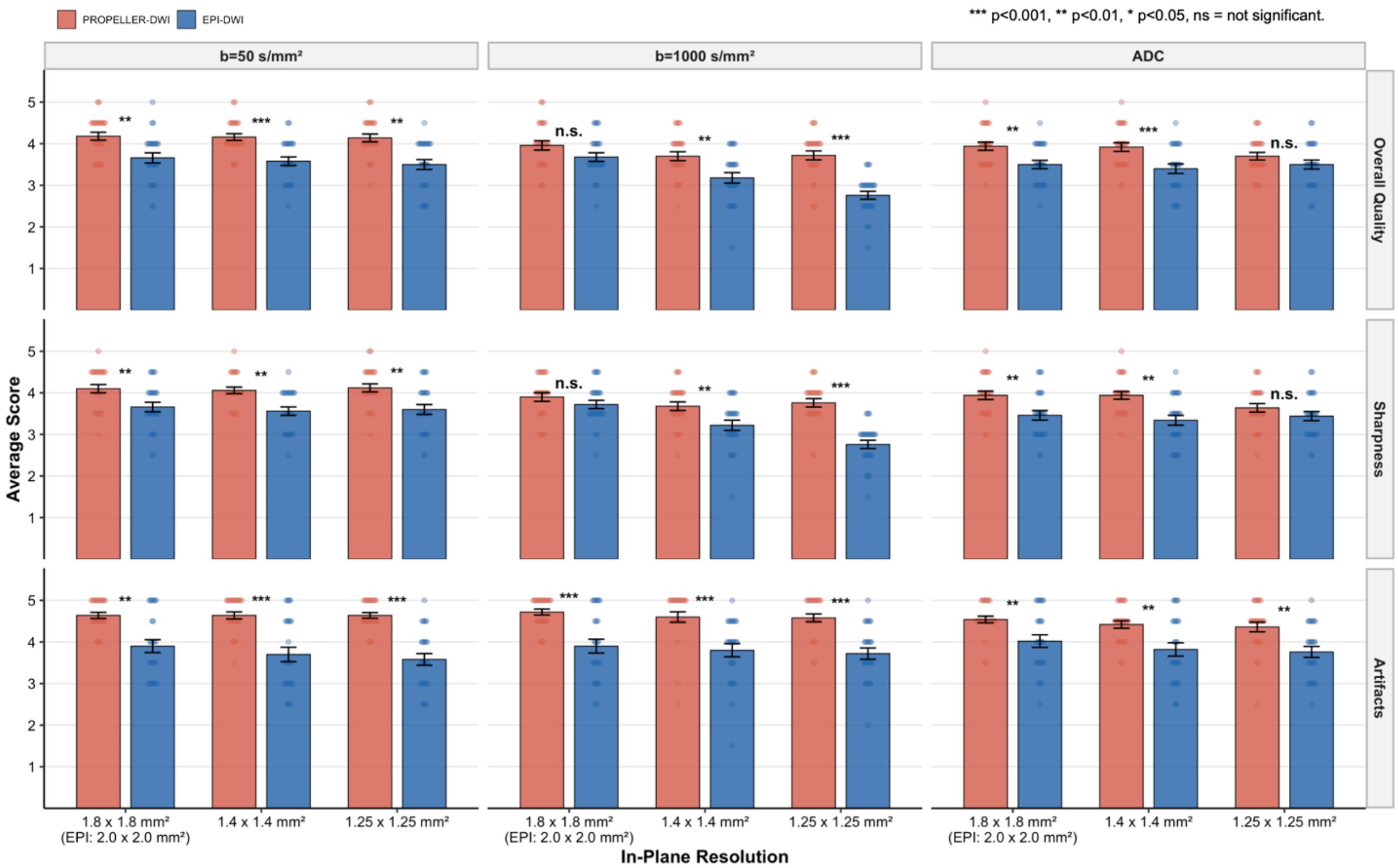

**Figure 8**. Reader image-quality assessment. Averaged reader scores for overall image quality, sharpness, and artifacts are shown for b=50 s/mm$^2$ images, b=1000 s/mm$^2$ images, and ADC maps across the three spatial-resolution settings. PROPELLER-DWI generally received higher scores than EPI-DWI, with the most consistent improvement observed for artifact scores. Error bars indicate standard error. Significance levels are indicated as *** p<0.001, ** p<0.01, * p<0.05, and ns = not significant.

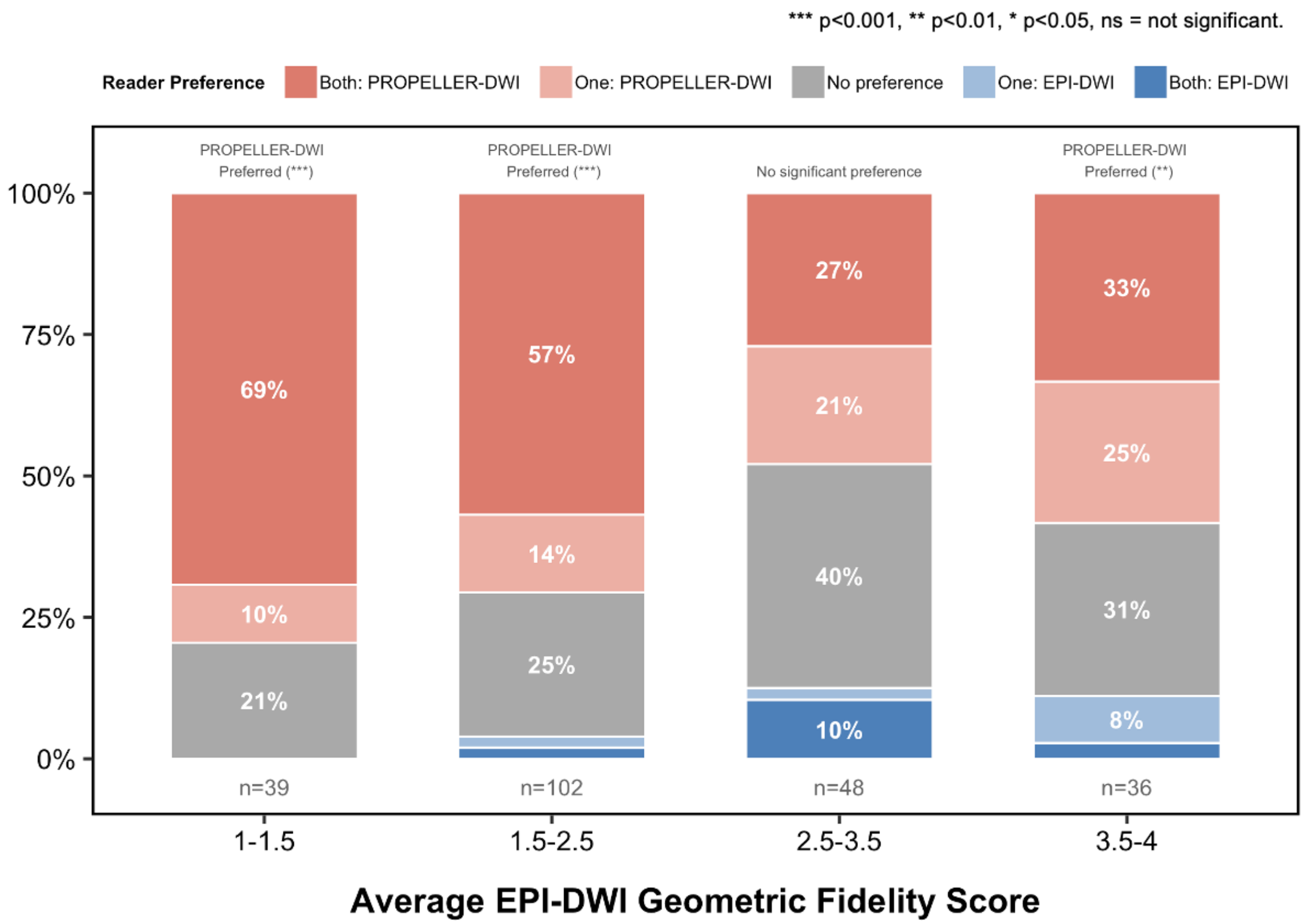

**Figure 9**. Reader preference stratified by EPI-DWI geometric fidelity. Reader preference for PROPELLER-DWI versus EPI-DWI was evaluated after grouping cases by the average EPI-DWI geometric fidelity score, with lower scores indicating greater visible distortion. Stacked bars show the proportion of cases in which both readers preferred PROPELLER-DWI, one reader preferred PROPELLER-DWI, neither reader had a preference, one reader preferred EPI-DWI, or both readers preferred EPI-DWI. The label above each bar indicates whether the overall preference within that bin was statistically significant by exact binomial testing. PROPELLER-DWI was significantly preferred when EPI-DWI geometric fidelity was poor to moderate and was also preferred in the highest-fidelity bin, whereas no significant preference was observed for the intermediate 2.5–3.5 score range. The corresponding resolution-stratified analysis is shown in Supporting Figure S6. Significance levels are indicated as *** p<0.001, ** p<0.01, * p<0.05, and ns = not significant.

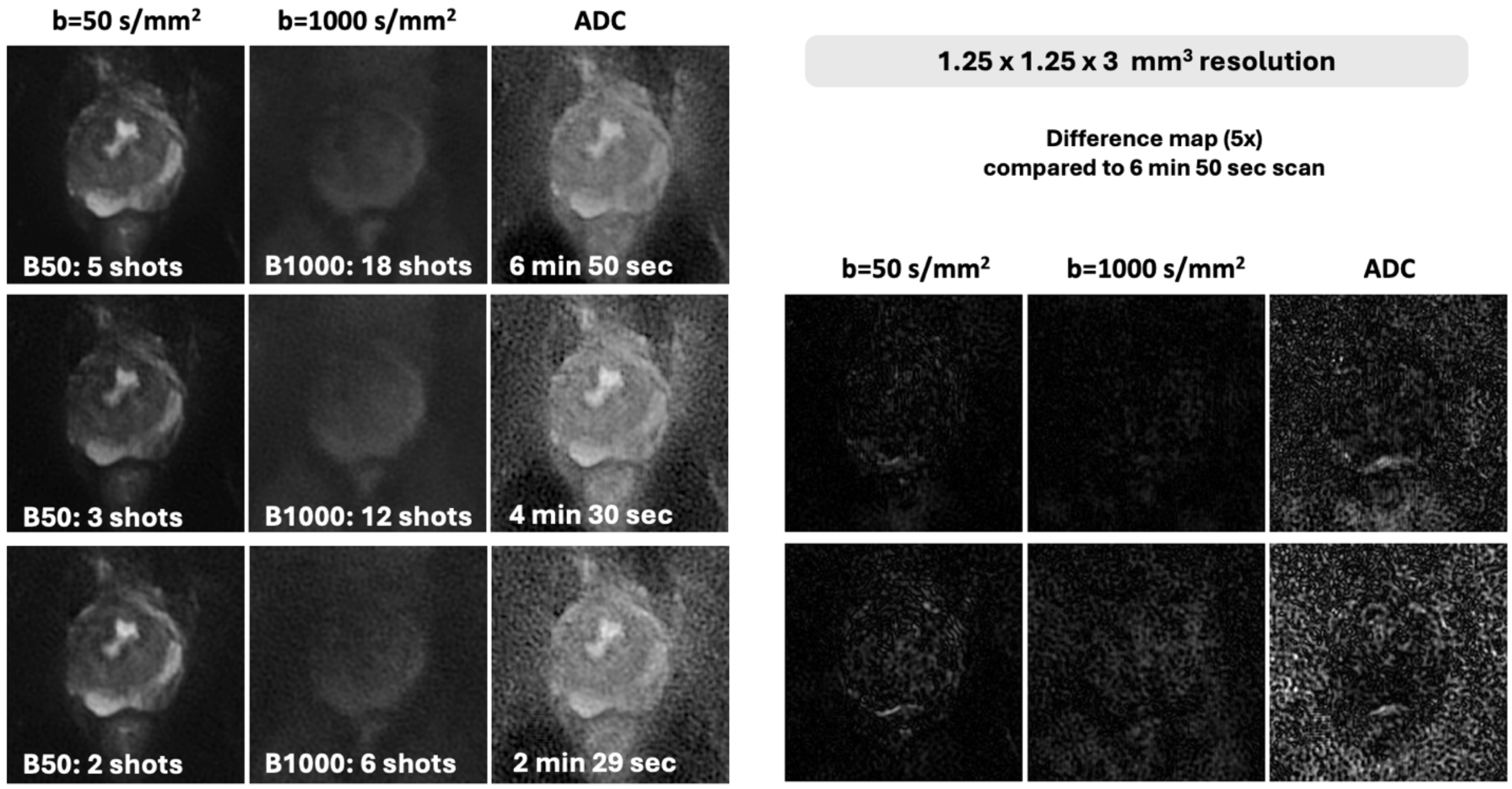

**Figure 10**. Retrospective reduced-shot PROPELLER-DWI reconstruction. Representative high-resolution PROPELLER-DWI images at 1.25 x 1.25 x 3 $mm^3$ resolution reconstructed using the full acquisition and retrospectively reduced shot numbers. The full acquisition used 5 shots for b=50 s/$mm^2$ and 18 shots for b=1000 s/$mm^2$, with a scan time of 6 min 50 s. Reduced-shot reconstructions used 3/12 and 2/6 shots for b=50/b=1000 s/$mm^2$, corresponding to scan times of 4 min 30 s and 2 min 29 s, respectively. Difference maps, displayed at 5x scale relative to the full acquisition, show increased noise and residual differences with greater shot reduction, while overall prostate anatomy and ADC appearance remain preserved.

**Supporting Table S1**. Imaging parameters for phantom and in vivo experiments. Parameters include sequence type, in-plane resolution, slice thickness, TE, TR, number of gradient echoes and spin echoes per shot, echo spacing for gradient echoes and spin echoes, number of shots or averages for each b-value, and total scan time.

**Supporting Table S2**. Five-fold cross-validation case assignments. Training and testing case numbers are listed for each cross-validation fold used to evaluate the self-supervised PROPELLER-DWI reconstruction. Each subject was reconstructed using a model trained without that subject, and a common representative subject was included across folds for model-consistency assessment.

**Supporting Figure S1**. Effect of the number of gradient echoes per refocusing on scan efficiency and artifact behavior. Two representative subjects are shown with axial T2-weighted images (Ax T2), single-shot EPI-DWI, and PROPELLER-DWI acquired with 1, 3, 4, 6, and 9 gradient echoes per refocusing while maintaining a fixed total of 36 blades. Yellow arrows indicate the distortion artifact. Increasing the number of gradient echoes shortened scan time (**Supporting Table S1**), but excessive gradient echoes produced residual blurring or smearing near stronger off-resonance sources, particularly metal. The six-gradient-echo configuration was selected for subsequent in vivo experiments based on the balance between scan efficiency and off-resonance artifact control.

**Supporting Figure S2**. Model-to-model consistency of self-supervised PROPELLER-DWI reconstruction. A representative dataset was reconstructed using all five independently trained cross-validation models at 1.8 x 1.8 x 3 $mm^3$, 1.4 x 1.4 x 3 $mm^3$, and 1.25 x 1.25 x 3 $mm^3$ resolution. Mean images, individual model outputs, and 5x-scaled difference maps are shown for b=50 s/$mm^2$ and b=1000 s/$mm^2$ images. Structural similarity index metric values were labeled on the images and remained high across models, with only small model-dependent differences, mainly in low-signal background regions.

**Supporting Figure S3**. Clinical add-on patient examples with PI-RADS 3 lesions. Two patient cases acquired as clinical add-on scans are shown with T2-weighted (T2w) images, clinical EPI-DWI ADC maps with and without annotation, and PROPELLER-DWI b=50 s/$mm^2$, b=1000 s/$mm^2$, and ADC images. Lesion location and ADC appearance on PROPELLER-DWI are concordant with the clinical EPI-DWI findings, supporting the feasibility of PROPELLER-DWI for lesion depiction in clinically relevant cases.

**Supporting Figure S4**. Left-right prostate ADC measurements. Mean ADC values measured from left and right peripheral-zone and transition-zone ROIs are shown for EPI-DWI and PROPELLER-DWI across the tested spatial-resolution settings. Paired measurements from each subject are indicated as gray dots and connected by gray lines. No significant left-right ADC differences were observed.

**Supporting Figure S5**. Geometric fidelity scores for EPI-DWI and PROPELLER-DWI. Average geometric fidelity scores from two readers are shown for EPI-DWI and PROPELLER-DWI across the three spatial-resolution settings. Higher scores indicate better geometric fidelity with less visible distortion, prostate shape deformation, or signal pile-up. PROPELLER-DWI demonstrated significantly higher geometric fidelity than EPI-DWI at all tested resolutions. Error bars indicate standard error. Significance levels are indicated as *** $p<0.001$, ** $p<0.01$, * $p<0.05$, and ns = not significant.

**Supporting Figure S6**. Reader preference for PROPELLER-DWI or EPI-DWI after stratifying cases by the average EPI-DWI geometric fidelity score at each spatial-resolution setting. Lower EPI-DWI geometric fidelity scores indicate more severe visible distortion. The label above each bar identifies the method favored by the readers and the accompanying significance level shows the result of the exact binomial test. As shown in the stacked bars, readers increasingly favored PROPELLER-DWI when EPI-DWI distortion was more pronounced. Reader preference also shifted toward PROPELLER-DWI at the highest spatial resolution, where EPI-DWI was more susceptible to distortion and signal loss. Significance levels are indicated as *** $p<0.001$, ** $p<0.01$, * $p<0.05$, and ns = not significant.

**Supporting Figure S7.** Comparison of PROPELLER-DWI, clinical single-shot EPI-DWI, and reduced-FOV acquisition ZOOMit in a subject with severe metal-induced susceptibility artifacts. Metal locations are indicated by red arrows on the axial T2-weighted images. Substantial distortion and signal loss were observed on EPI-DWI. ZOOMit reduced the spatial extent of the artifact, although residual distortion and signal degradation remained. Whereas PROPELLER-DWI better preserved the prostate

anatomy. The three-direction trace ZOOMit acquisition follows the clinical protocol with a TE of 70 ms, scan time of 6 min 38 s and a spatial resolution of 2 x 2 x 3 $mm^3$.